\documentclass[twocolumn,prb,preprintnumbers,amsmath,amssymb,floatfix]{revtex4} 

\usepackage{graphicx}
\usepackage{subfigure}
\usepackage{dcolumn}
\begin{document}

\title{The Next Breakthrough in Phonon-Mediated Superconductivity}
\author{W. E. Pickett}
\affiliation{Department of Physics, University of California, Davis, California, 95616}

\begin{abstract}
If history teaches us anything, it is that the next breakthrough in
superconductivity will not be the result of surveying the history of
past breakthroughs, as they have almost always been a 
matter of serendipity resulting from undirected
exploration into new materials.  Still, there is reason to
reflect on recent advances, work toward higher T$_c$ of even an
incremental nature, and recognize that it is important to explore
avenues currently believed to be unpromising even as we attempt 
to be rational.  In this paper we look at two remarkable new
unusually high temperature superconductors (UHTS), MgB$_2$ with
T$_c$=40 K and (in less detail)
high pressure Li with T$_c$=20 K, with the aim of
reducing their unexpected achievements to a simple and clear
understanding.  We also consider briefly other UHTS systems that
provide still unresolved puzzles; these materials include mostly
layered structures, and several with strongly bonded C-C or B-C
substructures.  What may be possible in phonon-coupled 
superconductivity is reconsidered in the light of the discussion. 
%C$_{60}$-based fullerides (T$_c$ up to 40 K);
%Ba$_{1-x}$K$_x$BiO$_3$ (T$_c$ up to 35 K);
%Li$_x$(Zr,Hf)NCl (T$_c$ up to
%25 K); Y$_2$C$_3$ (T$_c$=18 K). 

\end{abstract}
\maketitle

%\parindent 0pt
%\parskip 12pt

%%%%%%%%%%%%%%%%%%%%%%%%%%%%%%%%%%%%%%%%%%%%%%%%%%%%%%%%%%%
%  Introduction
\section{Motivation}
The appearance of several startling examples of superconductivity
in the past six year or so is prompting re-evaluation of our
thinking about the one pairing mechanism for which there is a
precise and controlled theoretical foundation, to wit, phonon-mediated
coupling. This strong-coupling Migdal-Eliashberg (ME) theory, 
formalized in 
detail by Scalapino, Schrieffer, and Wilkins,\cite{ssw} has had
numerous successes in the {\it quantitative} description of   
the frequency dependence of the complex superconducting gap
function, the deviation of the critical field from its weak-coupling
analytic form, etc.  Its implementation has not been so precise in
predicting the critical temperature T$_c$ because the retarded
Coulomb repulsion $\mu^*$ is difficult to calculate; nevertheless
the underlying theory of T$_c$ is understood to be in good
shape.

So a question arises: when we have a material-specific, quantitative
theory that works, why is it that we are nearly always surprised 
by the most interesting new cases of ``unusually high T$_c$
superconductivity''
(UHTS), which here we will consider to be around 20 K or higher.
(The high T$_c$ cuprates, the real HTS materials, are a separate
class and will not be considered here.)  In this paper we will
attempt to clarify issues that are involved in several of
these UHTS materials, and to illuminate some of issues in the 
{\it understanding} of the ME theory.  The aim of this paper is
to provide a generalized conceptualization of some of the new
surprises.  The aim is not, unfortunately, to predict the next
breakthrough, as the title misleadingly implies, although a
provocative limit will be mentioned.

One relevant system that will not be addressed here is the
fulleride superconductors A$_x$C$_{60}$, with values of T$_c$
up to 40 K being achieved.\cite{palstra}  
There is an enormous literature on this system, with many of the
important papers being cited in a recent review.\cite{olle}
It is rather unfortunate that we do not have the space-time nor the
energy to include fullerides, as some of its important characteristics
overlap strongly with those we discuss here.  There are however
correlation effects that complicate the theoretical description,
and therefore the comparison, with materials discussed here, and it
would not be prudent to draw parallels or contrasts in this paper.  

For different reasons no discussion of PuCoGa$_5$ 
(with its surprising T$_c$=18 K)\cite{sarrao} will
be included.  This heavy-fermion UHTS is a very different kind of
material than any that will be discussed in this article, and no
doubt requires a very different theoretical approach.

\section{MgB$_2$, the Queen of Phonon Coupling}
\subsection{Background}
The discovery by Akimitsu's group\cite{akimitsu} 
in 2001 of T$_c$ = 40 K in MgB$_2$ was unimaginable within the context
of conventional understanding at the time, as will be elaborated further
in this paper.  The measurement of the boron 
isotope shift\cite{canfield} quickly established a phonon mechanism,
and the structural and electronic simplicity allowed many groups to
dive into study of the mechanism.  The understanding of the mechanism
arose 
quickly\cite{jan,kortus,kong,bohnen,yildirim,liu,choi,extreme,2Dpickett,brazil}
and is in reasonable quantitative agreement with data. The truly 
remarkable aspect of this 
Queen of superconductivity's
personality traits is her complete and utter scorn for the
conventional wisdom of phonon-coupled superconductivity 
(``Matthias's rules''). 

{\it Broken rule \#1:} MgB$_2$ is not cubic nor is it close, 
and this is one of its key characteristics.  The all-important 
$\sigma$ band is 
(quasi) two-dimensional (2D), so although it is rather lightly
(self-)doped with holes, its density of states is comparable to  
what it would be with heavier doping.  In addition, the $sp^2$
bonding in the B graphene layer provides it with stronger bonds than
if it had three-dimensional (3D) $sp^3$ bonds.  This distinction in
$sp^2$ versus $sp^3$ bonding is why graphite is more strongly bound
than is diamond. 

{\it Broken rule \#2:} There are no $d$ electrons; previous emphasis
was on intermetallic compounds, {\it viz.} Nb$_3$Sn, 
where $d$ electrons played the
central role.  MgB$_2$ takes
advantage of the fact that $sp^2$ and $sp^3$ bonds are the strongest in nature,
stronger than the $d-d$ bonds in transition metal compounds that
had ruled the conventional superconductivity roost.  The strong
bonds lead to extremely large electron-phonon matrix elements; we
return to this below.

{\it Broken rule \#3:}  There is no special e/a ({\it electron/atom}) ratio
that tunes the Fermi level to a peak in the density of states N(E),
because N(E) has no peaks and furthermore its magnitude is embarrassingly modest.
MgB$_2$ exchanges large N(E) for very large matrix elements.  This
rule of large N(E$_F$) is justified by the expression 
for the coupling strength $\lambda$,
\begin{eqnarray}
\lambda = \frac{N(E_F) <I^2>}{M<\omega^2>},
\end{eqnarray} 
with the other quantities being the mean square electron-phonon
matrix element $<I^2>$ averaged over the Fermi surface, the ion mass
$M$, and the mean square renormalized (physical) phonon frequency.
(It will not be necessary for the purpose of this paper to delve
into the complexities that arise in compounds with more than a
single type of atom.)  Since T$_c$ increases monotonically with
$\lambda$ (all other characteristics kept fixed) and clearly
it is proportional to N(E$_F$), then a higher density of states
is desirable.  [We point out below the incorrectness of this argument;
$<\omega^2>$ also depends on N(E$_F$).]

{\it The rule of light elements.}  This rule was not included in the
conventional list, because it was not clear there was any real correlation
between T$_c$ and mass in the best intermetallic superconductors;
Nb$_3$Sn was as good a Nb$_3$Al, for example, and a little better
than the much lighter V$_3$Si.  Theoretically, however, it was accepted
that having a high energy boson doing the coupling [T$_c \sim 
\omega_{boson}$exp(-1/$\lambda$)] provides a higher
energy scale, and therefore lends hope for driving T$_c$ skyward.  Metallic
hydrogen is the limiting case (barring the formation of a
condensed system of muonium atoms, for which the $\mu^+$ lifetime
becomes an issue).  Ashcroft's prediction 35 years ago that it would
be a high temperature superconductor\cite{ashcroft} remains
unverified, but also remains unretracted.  A recent compilation of
the values of T$_c$ for elemental metals\cite{buzea} shows the high
values to be concentrated toward the low mass end of the spectrum.

A more microscopically based overview of the interrelationships between
large susceptibilities [N(E$_F$), $\chi(q)$], large matrix elements,
strong interatomic forces, and atomic masses was provided by 
Allen,\cite{allen80a,allen80b,allen82} who also 
focused on the limitations posed by structural
instabilities as coupling became too strong.
What is sobering to recognize is that the behavior of MgB$_2$ lies
within conventional Migdal-Eliashberg theory; it was only our
biases (as codified in Matthias's rules) that MgB$_2$ abused.
What MgB$_2$ really did in spectacular fashion was to violate 
a rule we probably
were often not consciously aware that 
we followed (although we certainly ``understood'' it).\\

{\it Broken unwritten rule A.} The unwritten rule, the ``eleventh
commandment'' of successful electron-phonon systems, can be
phrased as {\it thou shalt not put all of thine eggs into one basket}.
Successful electron-phonon coupling system should refrain from being too
pushy, the coupling had to be spread out over most
(preferably all) phonons, {\it i.e.} many baskets.   
Extremely strong coupling to any given phonon was the recipe for
banishment from superconducting materials, via structural instability.   
The decomposition of the coupling strength $\lambda$ into contributions
from individual phonons (mode $\lambda$'s, $\lambda_Q$) 
by Allen\cite{allen} made
the connection clearer, and has also been the key to understanding
coupling strength in the UHTSs.
This {\it many-baskets} rule was not so 
clearly codified but nevertheless
was quite clear: there were several examples where coupling strength
got unusually strong at certain values (or localized regions) of phonon
wavevector Q.  The phonon branch softened, as parameters were
twiddled to increase the coupling the phonon became unstable, and
this enhanced T$_c$.  Certainly these softened branches correlated 
closely with increased
T$_c$, and there was a manifestation of the increased strength of
the coupling.  The first instance was probably the TaC-HfC pair
measured by Smith and Glaser\cite{glaser}, and was followed by the
Nb$_3$Sn-Nb$_3$Sb distinction.\cite{pintschovius}  
The theory provided understanding of these connections,
and also the means\cite{weppba} to 
obtain the enhancement of the coupling (as long as anharmonic corrections
were not important).  If the renormalization became too strong, however,
the compound transformed to a structure with weaker coupling (Peierls-type
distortion due to strong Kohn anomaly, or a band Jahn-Teller transformation
if N(E$_F$) became too large), or the structure
would not form at all (``covalent instability'').

\subsection{The Secrets of MgB$_2$}
MgB$_2$ made, to put it kindly, fools of those of us who believed that
focusing of coupling strength into a few modes was folly.  [The author
was a fervent believer in this `evident truth.']  Only two of the nine
phonon branches are strongly coupled, these being the doubly degenerate
B-B bond-stretching modes in the 2D layers.  And of these branches,
only those with $Q < 2k_F$ are strongly coupled, and this set comprises only
12\% of the area of the zone.  Thus only 3\% of the MgB$_2$'s phonons 
are carrying the load, with ``mode $\lambda_Q$'' values of $\sim$20-25; 
the other 97\% have values two orders of magnitude smaller and serve
mainly to complicate the analysis and confuse the understanding, which
is actually exceedingly simple.  Neglecting these (97\%) dismal wannabes, a
very good estimate of the coupling strength can be obtained with a
handful of easy computations and the back of a clean standard-size
envelope.\cite{brazil}

The explanation is most easily visible in the calculation of the phonon
spectrum of hole-doped LiBC, whose crystal and electronic structure 
is much like those of MgB$_2$ (as are the predicted phonon
anomalies\cite{LiBC})
except the matrix elements are even larger.  The phonon dispersion curves,
before and after hole-doping, are shown in Fig. 1.  The 
enormous Kohn anomalies are sharper than those published for MgB$_2$,
possibly because it is a little more 2D-like, but more so because the
phonon momentum grid used in the calculation is much denser than has
been done for MgB$_2$.  Just as simple considerations based on the
circular Fermi surface suggest, renormalization is confined to $Q<2k_F$,
but the strength of renormalization is unprecedented.  The extreme phonon
softening and broadening, and the sharp Kohn anomalies, have been
experimentally verified in MgB$_2$.\cite{quilty,martinho,shukla}

Predictions of the sort
that give realistic values for MgB$_2$ suggest that in Li$_x$BC, T$_c$
could be as much as twice as high.\cite{LiBC,dewhurst}
Li$_x$BC was reported a decade ago by
Worl\"e {\it at al.},\cite{worle} but was not checked for 
superconductivity.  Recent attempts at Li de-intercalation have either not been 
successful.\cite{bharathi,hlinka,renker,souptel,pronin} or 
have not produced a metallic material.\cite{zhao,fogg2}

\begin{figure}[t]
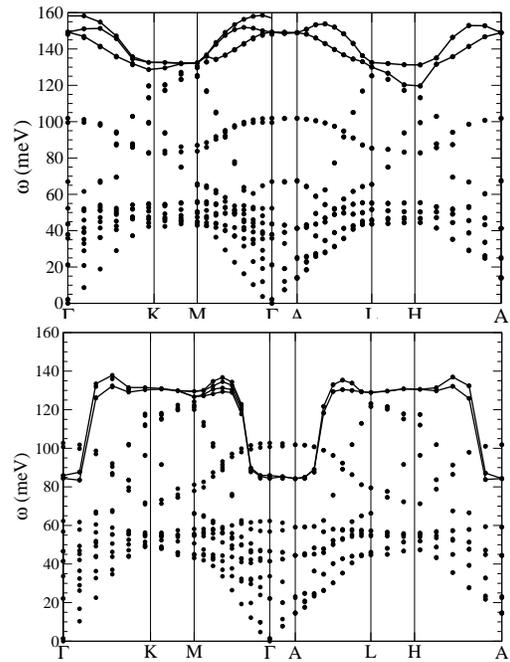

\label{libc}
\includegraphics[width=0.40\textwidth]{LiBC.wq.eps}
\includegraphics[width=0.40\textwidth]{LixBC.wq.eps}
\caption{Calculated phonon dispersion curves for the MgB$_2$ spinoff
Li$_x$BC.  Top panel: phonons for the semiconducting $x=1$ compound,
with the B-C bond-stretching modes emphasized by connecting the
calculated points.  Bottom panel: corresponding phonons for 25\%
hole doping ($x$=0.75).  The Kohn anomaly at $Q=2k_F$ is like
the one in MgB$_2$ but more extreme.  It is also a sharper drop at
$2k_F$ than published results for MgB$_2$ because the phonons were
calculated on a denser mesh.  The relation to the model calculation
in Fig. 2 is clear.
}
\end{figure}
\begin{figure}[t]
\label{vsKF}
\includegraphics[width=0.40\textwidth]{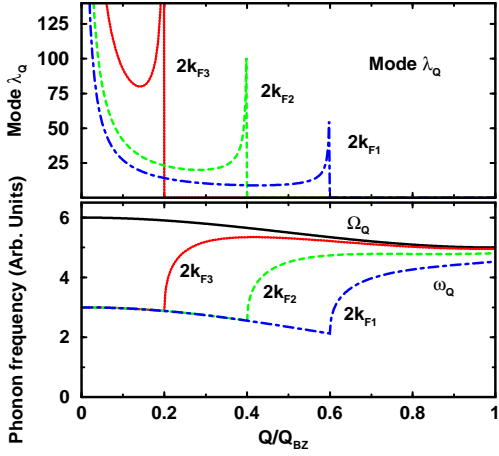}
\caption{Results of circular Fermi surface model showing the change
in the mode $\lambda_Q$ (top panel) and the renormalization of the
bond-stretching mode (bottom panel).  Three values of $2k_F$ are
pictured; $2k_F$ increases with additional hole doping.  Features
to note: values of the mode $\lambda_Q$ decrease with increased doping,
but more modes have the large mode $\lambda_Q$, and as a result the
total $\lambda$ in independent of doping level; the amount of
renormalization (bottom panel) is independent of the doping level,
even though more phonons get renormalized as $2k_F$ increases.
}
\end{figure}

\subsection{Extrapolating from MgB$_2$}
Staying within the MgB$_2$ paradigm, one can ask the perverse question:
why is T$_c$ of MgB$_2$ {\bf only} 40 K? why isn't it 60 K, or 100 K?  Is it
within the realm of possibility that an MgB$_2$-type material could be   
a room temperature superconductor?  It is worthwhile to pursue this line
of reasoning, neglecting for the time being that the biggest lesson that
history has taught us is that every qualitative jump in T$_c$ does not
result from scientific scheming but, maddeningly, simply from serendipity.
[Examples: high T$_c$ cuprates; fullerides; MgB$_2$.  In none of these
systems was the remarkable superconductivity foreseen.  The exception:
T$_c$ up to 35 K in (Ba,K)BiO$_3$ was actually the outgrowth of
systematic scientific scheming at Bell Labs by Mattheiss and
coworkers.\cite{mattheiss}]

\begin{figure}[t]
\label{trends}
\includegraphics[width=0.40\textwidth]{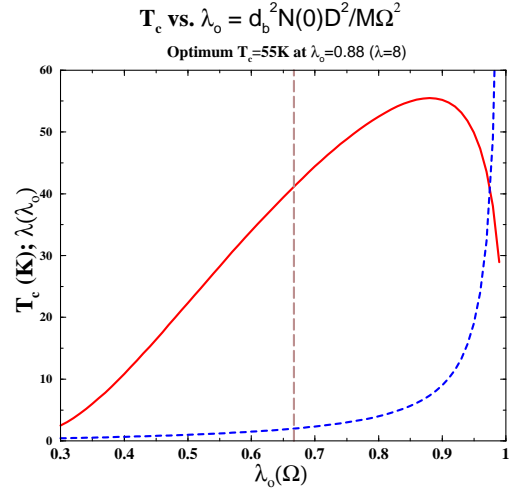}
\includegraphics[width=0.40\textwidth]{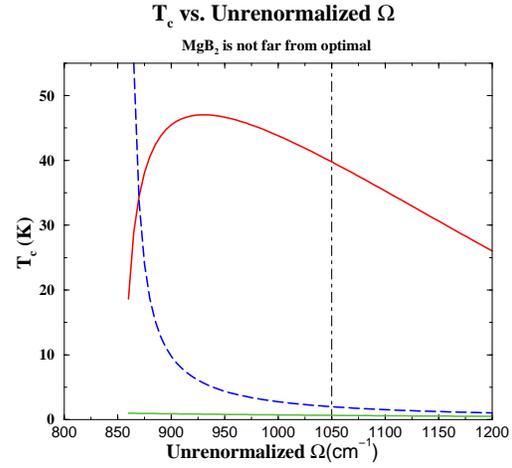}
\caption{Change in T$_c$ with variation of chosen characteristics
of MgB$_2$.
}
\end{figure}

Fortunately, MgB$_2$ did not break the essential rule for 
good electron-phonon coupled
superconductors: that Migdal-Eliashberg theory provides the description
of not only T$_c$ but the wavevector and frequency dependence of the
superconducting gap.  The underlying theory is still the one we understand,
so one can pursue the theoretical game of varying specific
materials characteristics individually, do learn what their influence is.

One of the first issues to consider, and one not directly related to
Matthias's rules, is the doping dependence of $\lambda$ and T$_c$.  This
dependence arises mainly from the scale $k_F$ of the Fermi surface
(we neglect the renormalization of interatomic forces and bands).  
An analytic
(front of the envelope) calculation leads to the results pictured in Fig.
2 and the remarkable implications.  The coupling 
strength $\lambda_Q$ is confined to $Q < 2k_F$ and
decreases as $k_F$ (hole doping) increases.  
However, the phase space, that is, the
Fermi surface volume, increases in exactly a manner that leaves $\lambda$
itself constant: the total coupling strength is independent of the
doping level.  The next result, directly following, is that the 
phonon renormalization does not change, even though an increasing 
fraction of the phonons are renormalized.  Thus simple doping changes 
the number of phonons that are renormalized (following the 
{\it many-baskets} theme) but the mode $\lambda$'s are decreased,
so doping is ineffective in effecting an increase in
T$_c$ in a system like MgB$_2$.

%\begin{eqnarray}
%T_c &=& \frac{f_1 f_2 \omega_{log}}{1.20}
% e^{-\frac{1.04(1+\lambda)}{\lambda - \mu^* -0.62 \lambda \mu^*}},\\
%f_1 &=& [1+(\lambda/\Lambda_1)^{3/2}]^{1/3},\\
%f_2 &=& 1+\frac{(\omega_{2}/\omega_{log}-1)\lambda^2}{\lambda^2 + \Lambda_2^2},\\
%\Lambda_1 &=& 2.46(1+3.8\mu^*),\\
%\Lambda_2 &=& 1.82(1+6.3\mu^*) (\omega_2/\omega_{log}).
%\label{AllDyn}
%\end{eqnarray}
%This form was constructed to retain the exponential expression involving 
%$lambda$ and the retarded Coulomb repulsion $\mu^*$.  
%The strong-coupling
%corrections involve in addition the logarithmic and second frequency moments 
%$\omega_{log}$ and $\omega_2$ respectively.  It is the important $f_1$ factor
%that results in the correct result $T_c \propto \sqrt{\lambda}$ at large
%coupling.

The first broken rule, about symmetry of the crystal structure, cannot be  
fixed in a continuous way, since the hexagonal structure of MgB$_2$ cannot
be morphed into a cubic counterpart in any way that provide a useful 
comparison (although a quasi-2D electronic structure may be connected
continuously to a 3D one).  Likewise, $p$ electrons cannot be 
squeezed continuously into
$d$ electrons, again a question of symmetry.  The Queen has violated these two
rules in a nonnegotiable fashion, as in a royal decree.

The third rule does involve a variable quantity, N(E$_F$), which 
{\it can} be changed by varying the effective mass $m^*$ or by encountering
a non-parabolic dispersion relation.  However, $<I^2>$ enters the equations
multiplied by N(E$_F$) so they can be considered together.  It gets
better than that: the unrenormalized phonon frequency $\Omega$ enters 
similarly, except in the form of its inverse square.  Simply put, once
phonon renormalization in this 2D system has been incorporated, one 
can express\cite{jan,kong,extreme}
$\lambda$ in terms of the intrinsic material parameters encapsulated in
$\lambda_o$:
\begin{eqnarray}
\lambda=\frac{\lambda_o}{1-\lambda_o}; \lambda_o = 
  \frac{d_B^2 N(E_F) <I^2>}{M\Omega^2}.
\label{lambda}
\end{eqnarray}
Here $d_B$ is the band (Fermi surface) degeneracy; for MgB$_2$ there are
two Fermi surfaces $d_B = 2$ and a reward of $d_B^2$=4.

With this form we can consider the variation of $\lambda_o$ to be due
to variation of N(E$_F$), $<I^2>$, or $d_B$.  Keeping in mind the
frequency prefactor in the equation for T$_c$, which is the renormalized
frequency (and depends on $\lambda_o$ in a known way), the change in
T$_c$ can be obtained.
For numerical realism in the strong coupling regime we are probing,
we must use instead of the McMillan equation
the Allen-Dynes equation,\cite{allendynes}
which gives the correct strong-coupling limit
$T_c \propto \sqrt{\lambda}$.  For the results we present in Eq. 3,
$\mu^*$ has been adjusted to provide the reference value T$_c$ = 40 K
for MgB$_2$ when first-principles results are used for the other
quantities. [The AlB$_2$ phonon frequency $\Omega$ = 1050 cm$^{-1}$
was taken as the unrenormalized frequency.]

The top panel of Fig. 3 shows, in addition to the trivial but
instructive behavior of $\lambda(\lambda_o)$, the variation of T$_c$
as $d_B^2$ N(E$_F$) $<I^2>$ is varied, {\it i.e.} varying $\lambda_o$
at fixed $\Omega$.  It is seen that a higher band
mass, or larger matrix element, can increase T$_c$ by only $\sim$30\% 
to 55 K (where $\lambda \sim$ 7-8);
at this point the renormalized frequency crashes toward zero and
in spite of a divergent $\lambda$ (achieved by a vanishing denominator
$M<\omega^2>$) T$_c$ drops to zero.  The other variation to consider
is that of keeping the electronic characteristics fixed by varying the
unrenormalized frequency.  It is found that reducing $\Omega$ initially
leads to an increase in T$_c$, but after only a 15\% increase at
around $\Omega \sim$ 900 cm$^{-1}$ (again, $\lambda \sim 7-8$), 
at which point the renormalized frequency again
comes crashing down and T$_c$ vanishes.  Strictly, it seems in both
cases that T$_c$ does not vanish before the system becomes unstable.
What we see is that our old conventional picture finally wins out:
increased coupling strength leads to instability after only a rather
modest enhancement of T$_c$ for the case of MgB$_2$.  The Queen
loses her head after all. 

So, is there no way to win, no scenario that gives room temperature
superconductivity?  The lesson of Fig. 3 really is that the 
instability limit $\lambda_o \rightarrow$ 1 must be avoided, and at 
least in these scenarios the ultra-strong coupling
T$_c \rightarrow \sqrt{\lambda}$ scaling,\cite{allendynes}
derived for fixed frequencies, is not part of the parameter space.
However, if $\lambda_o$ can be kept near its optimum value of 0.9
and the frequency (prefactor) increased, then there is no limitation
on the increase.  This situation might be obtained in two ways.  One
is to increase the numerator and denominator of $\lambda_o$
proportionately; make the electronic coupling stronger while also
making the underlying (unrenormalized) lattice stiffer.  The increase
in T$_c$ then follows the increase in the frequency prefactor, which
although renormalized is still increasing.  MgB$_2$, and B-doped
diamond even more so, capitalize on a stiff underlying lattice, but 
hard lattices are limited at around the diamond example.  The 
alternative is to apply pressure, and indeed pressure is seen to
enhance T$_c$ impressively in many systems.

The other possibility, based on the fact that $M<\omega^2>$ is actually
a force constant and is mass independent (hence $\lambda_o$ is 
mass independent), is to fix all material parameters and simply reduce
the nuclear masses.  This is really the ``metallic hydrogen'' 
limit mentioned above, and such variation of masses via isotope
substitution is quite limited in practice.   

\subsection{Boron-doped Diamond: MgB$_2$ in 3D}
Although B-doped diamond does not (yet) fit our $\sim$20 K criterion
as a UHTS, it is quite instructive to consider it because of its
relationship to MgB$_2$.  Diamond, doped at the 2-4\% level by boron,
has been shown\cite{ekimov1,takano,bustarret,ekimov2,umezawa,takano2}
to be superconducting up to 11 K in bulk or thin film form.  Application
of ME theory, presuming that the material is a degenerate p-type
semiconductor, shows\cite{boeri,ucd1,blase,xiang,ma} strong 
electron-phonon coupling of a magnitude
that will account for its observed T$_c$.  ARPES data has verified that
the Fermi level indeed lies within the diamond valence bands at the
expected energy\cite{arpes} rather than in an impurity band as has
been speculated.\cite{pogorelov,baskaran}

The calculations show this system to be a 3D analog of MgB$_2$; indeed
one of the papers is so titled.\cite{boeri}  Holes are doped into 
the strongly bonding states, which are very strongly coupled to the
C-C bond stretch modes -- just the story of the high T$_c$ in MgB$_2$.
In terms of the quantities involved in the coupling of MgB$_2$ (above),
the comparison is the following.  The unrenormalized frequency, the
1330 cm$^{-1}$ mode of diamond, is higher than its MgB$_2$ analog (the
Raman mode of AlB$_2$ at 1050 cm$^{-1}$; thus the lattice is 
stiffer.\cite{cardona}
The $<I^2>$ matrix elements are larger than in MgB$_2$, due again to
the shorter stronger C-C bond compared to B-B.  Yet the renormalized
frequency, $\omega \approx$ 1000 cm$^{-1}$ remains higher than that of
MgB$_2$; this is the prefactor in the T$_c$ equation and also is good. 
The only shortcoming, and a severe one, is that diamond is 3D, which
means that the N(E)$\propto \sqrt{E_o - E}$ increases slowly with doping
below the band edge $E_o$.  As a result N(E$_F$) is much smaller than 
in MgB$_2$, by about a factor of four.

\section{Lithium Under High Pressure}
The unexpected superconductivity of MgB$_2$ is perhaps matched by the
subsequent discovery that the free-electron metal Li, upon being 
subjected to 35-50 GPa pressure, corresponding to a volume of only
40-50\% of its zero pressure value, becomes 
superconducting\cite{shimizu,struzhkin,schilling} at up to
20 K.  Since the upper limit of T$_c$ in Li at ambient pressure has been
decreased\cite{fins} to 100 $\mu$K, 
this increase represents at least a five order of
magnitude due to pressure.   This 20 K value
gives Li the
highest T$_c$ among elemental superconductors.  How can it happen that
a simple $s$-electron metal, still in a simple close-packed structure 
(fcc) suggestive of conventional metallic bonding, can produce 
the strength of electron-phonon
coupling that is necessary, as two studies now are 
showing.\cite{LiUCD,LiGross}

Those papers can be consulted for the details, but the basic physics 
goes like this.   Reduction of the atomic volume by a factor of
$\sim$2 doesn't produce enormous changes in the band structure; the
new, and crucial, feature is the appearance and growth in size of necks joining
spherical Fermi surfaces along $<111>$ directions, as for the well
known Fermi surface of Cu.  There is also flattening (with respect to
spherical) of the Fermi surfaces between the necks that has some
import; see below. These necks also develop primarily $2p$
character, thus Li is transforming from an $s$ electron metal into an
$s-p$ metal.  This $p$ character provides at least the possibility of
some directional, `covalent' character to the bonding, with the possibility
that this may enhance electron-phonon coupling.  

With the increase of pressure in the 20-38 GPa range where the fcc phase
is stable, the transverse T$_1$ branch of the phonon spectrum (and
only this branch) softens and becomes harmonically unstable around
25-30 GPa,\cite{LiUCD,LiGross} reflecting the 
renormalization that results from strong 
electron-phonon coupling.  Searching through the phonon scattering
processes from/to the Fermi surface, which is quantified by the 
``nesting function''
\begin{eqnarray}
\xi(Q) = \sum_k \delta(\varepsilon_k)\delta(\varepsilon_{k+Q}),
\label{xiQ}
\end{eqnarray}
reveals that the fairly innocuous
looking Fermi surface geometry actually focuses the phase space for
scattering processes into a few regions,\cite{LiUCD} 
including the one (Q near the
zone boundary K point) where the phonon branch becomes unstable. The
regions where the mode $\lambda_Q >$ 5 have also been mapped,\cite{LiGross}
and they appear to coincide with the regions of intensity in $\xi(Q)$.  

What has this enormous, and unanticipated, increase in T$_c$ with pressure
in Li to tell us about the bigger picture?  (1) Li is cubic here, with 
the simplest of structures, so there is no low-dimensionality effect
operating here. (2) Li is an $s-p$ electron material at reduced volume,
a quality in common with MgB$_2$.  (3) The electron-phonon coupling
strength is {\it not} driven by a large N(E$_F$), again similar to MgB$_2$.
(4) Very much like MgB$_2$, Li obtains its coupling strength from a
relatively small fraction of the phonons: the region around 
$Q=(2/3,2/3,0)2\pi/a$ has strongest coupling, but only to phonons with
transverse polarization $<1\bar{1}0>$.  In the case of Li, however, 
the sharp structure in Q of the coupling strength does drive the system to structural 
instability, at least in harmonic approximation.  Thus
the mechanism driving the increase in T$_c$ is 
self-limiting already
at experimental conditions as the conventional 
theory\cite{allen80a,allen80b,allen82} would lead one to expect.
Current data suggest, however, that the high T$_c$ survives the 
structural change at 40 GPa, and displays somewhat higher T$_c$
in the higher pressure phase.\cite{shimizu}

\section{Outstanding Puzzles in EP Superconductivity}
In this section we mention additional UHTS materials with T$_c \sim$ 20 K or
higher whose origin is unexplained.

\subsection{Electron-doped Hf \& Zr Nitridochlorides}
Electron doping (for example by intercalating Li) of HfNCl leads
to T$_c$ = 25 K; for Li$_x$ZrNCl the value is 12-15 K; these values depend
only weakly on doping.\cite{zrncl,sham1,hfncl,cros}   
The structure consists
of graphene-like honeycomb double sheets of alternating Zr and N, in such a 
way that each atom is bonded with three of the other type within a layer, and
with one of the other type in the neighboring layer.  
Layers of Cl on either side of
this double layer results in an insulating  slab that is 
formally Zr$^{4+}$N$^{3-}$Cl$^-$.
These closed shell sheets are van der Walls bonded in the undoped
material, and Li intercalation leads to electron doping -- one
electron per Li -- and 2D metallicity and superconductivity.  
There is considerable covalence\cite{weht} to the Zr-N bonding, 
however, so this
cannot be pictured simply as an ionic system.

\begin{figure}[t]
\label{zrnclBS}
\includegraphics[width=0.40\textwidth]{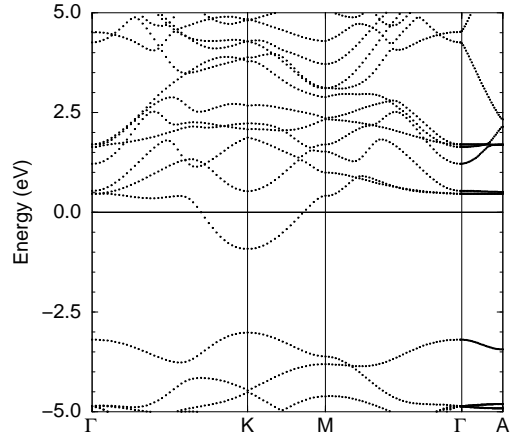}
\caption{Band structure of the electron-doped layered nitride 
Na$_{0.25}$ZrNCl, showing the single light-mass band into which
the electrons are doped.  The band minimum, and center of the
resulting Fermi surfaces, lie at the zone corner symmetry points K.
Note from the lack of dispersion along $\Gamma$-Z the strong
two-dimensionality of the electronic structure around the Fermi
level.
}
\end{figure}

The electron is doped into a small mass band [light carriers, low N(E)],
shown in Fig. 4,
with primarily Zn $d_{xy}, d_{x^2-y^2}$ in-plane character and
some N $p_x, p_y$ involvement.\cite{weht,felser,weht2,hase}  
There are circular
Fermi surfaces at the zone boundary K symmetry points.  A full calculation
of the phonon dispersion and electron-phonon coupling spectrum by
Heid {\it et al.} concludes\cite{heid} that the coupling strength 
$\lambda = 0.5$ is 
insufficient to account for T$_c$=15 K in Li$_{1/6}$ZrNCl.  The situation
is clearly analogous to MgB$_2$:  2D electronic system, circular Fermi
surfaces.  There being three (symmetry related but distinct) Fermi
surfaces, there are two sorts of Fermi surface scattering processes:
intrasurface, with $Q < 2k_F$; and intersurface, with $|Q-K| 
< 2k_F$.  Note that here
the point $\vec K$ arises because the centers of the Fermi surfaces are
separated by $Q\equiv K$, not because they also happen to be located at $K$. 
There are 3 intrasurface scatterings; there are 3! intersurface processes,
for a total factor of 3$^2$=9 similar contributions.
Each has the same circular phase space factor $\xi(Q)$ discussed above 
for MgB$_2$; the intra- and inter-sheet matrix elements may differ, however.  
The point is
that, whereas MgB$_2$ with its two Fermi surfaces takes 
advantage of the 2$^2$=4
factor, in these nitridochlorides there is a more robust 3$^2$=9 degeneracy 
factor.  Unfortunately, the matrix elements and/or N(E$_F$) magnitudes
are not sufficient to take advantage of this degeneracy.\cite{heid}

The Zr$\leftrightarrow$Hf comparison reveals a conundrum, which appears
to be separate but could be the crucial clue.  Why does the Hf system show
T$_c$ = 25-26 K, while the Zr materials are all T$_c$ = 15 K or a little
less?  Chemically, these isovalent $4d$ and $5d$ atoms are similar; in
any case, the larger atom never gives a stiffer lattice so that tendency
is backward.  Regarding them as chemically equivalent, the difference
could be viewed as an isotope shift (with masses differing by a factor
of two).  Again, the tendency is in the wrong direction, as it
is the (twice as) heavy element which has the (60\%) larger value of
T$_c$.  The simplest interpretation is that ME theory isn't working here,
and other mechanisms must be considered.  Bill and coworkers have
suggested\cite{bill} that an electronic mechanism may be operating;
still there is the Zr$\leftrightarrow$Hf question to address.

\subsection{BKBO: A Case unto Itself}
Ba$_{1-x}$K$_x$BiO$_3$ (BKBO), with T$_c$ up to 35 K reported,\cite{bkbo2}
has been relatively heavily studied with no resolution of the source
of its impressive superconductivity.  The parent compound BaBiO$_3$ is
a distorted perovskite with two inequivalent Bi sites, often 
interpreted formally as Bi$^{3+}$ (6$s^2$ ``lone pair''
configuration) and Bi$^{5+}$
(closed shell ion).  Extensive electronic structure studies find
very little actual charge difference between the sites, but these same
local density functional
studies are also unable to describe the ground state structural properties
as well as can usually be done for an $s-p$ electron system (the 
valence-conduction orbitals are O $2p$ and Bi $6s$).  

A density functional linear response calculation of the phonon spectrum,
electron-phonon coupling, and T$_c$ by Meregalli and Savrasov\cite{meregalli}
failed to find an explanation of the superconductivity.  The coupling
strength $\lambda$=0.34 was found, well below the required strength.  
Possible resolutions include (1) electron correlation, although how to
approach it is an open question (see below), (2) unconventional types of coupling to
the lattice; after all, this system superconducts best near a structural
phase boundary, (3) alloy issues (clustering or other short-range
order), which could make the virtual crystal treatment of the electronic
structure inapplicable, or (4) focusing of phonon scattering processes
as in Li, which makes it numerically taxing to carry out the Q-integration
to convergence to obtain the precise value of $\lambda$.
The rounded cube Fermi surface does suggest nesting that could make 
alternative (4) worth revisiting.

One course of action is to look for correlations beyond LDA
in this system.  The focus is on the Bi cation and its tendency to favor
the singlet-paired $6s^2$ configuration if electrons are available,
and the lack of any magnetic behavior.
This has been addressed in terms of a ``negative U''
interaction\cite{yoshioka,varma}
on the Bi ion, but a microscopic investigation into this possibility
failed to turn up evidence for such interaction.\cite{weber}

\subsection{Assorted UHTSs: Any Rhyme or Reason?}
There are a few other systems which fall under our classification as
UHTS materials, but are as yet little understood.

{\bf C$_2$ Dumbbell Systems.} The cubic compound Y$_2$C$_3$, 
consisting of a bcc structure
with Y occupying rather low symmetry $(u,u,u)$ sites, and with
interstitial dumbbells of C$_2$ molecules oriented along each of
the cubic axes, 
superconducts\cite{y2c3A,y2c3B} at 18 K when synthesized at high
pressure.  Keeping in mind MgB$_2$ and B-doped diamond (and fulleride
systems) with their strong coupling to high frequency B/C modes, it is
natural to focus on the unusual C$_2$ dumbbells, which are essentially
triple-bonded carbon molecules lying in the background electron gas provided
by the Y carriers.   Band structure calculations\cite{djsiim,shein} 
indicate a modest value of
N(E$_F$), but with an intriguing flat band very close to E$_F$ in
a limited region of the Brillouin zone.  Singh and Mazin calculate\cite{djsiim}
a C-C stretch mode frequency of 1442 cm$^{-1}$, higher than that of
diamond.  Its mode $\lambda_Q$ was only
about 10\% of that from the Y mode they studied.  It must be kept in
mind, however, that coupling to a high frequency mode is more valuable for
T$_c$ than coupling to a soft mode.  While it is 
suggested\cite{djsiim,shein}
that phonons may provide the coupling in this system, the most
distinctive feature of the structure, the C-C molecular dumbbells,
does not seem to be a dominant force in the impressive value of T$_c$.
On the other hand, the existing information is for only two phonons
at a single Q value in a system with 60 branches, and recent experience
teaches that as little as a few percent of the modes may drive T$_c$ up
to 30-40 K.  Unfortunately,
with 20 atoms in the primitive cell it is unlikely that full
electron-phonon coupling calculations can be carried out in the
near future.  

Another dumbbell system is the class
Y$_2$C$_2$${\cal H}$$_2$, where ${\cal H}$ is a negative halide ion 
(I, Br), with T$_c$ up\cite{simon,henn} to 11.5 K.   As for the Y$_2$C$_3$
systems, a C-C derived band lies very close\cite{puschnig} 
to the Fermi level.  Again
as for Y$_2$C$_3$, it is unknown whether this band contributes strong
electron-phonon coupling that drives the superconductivity in this
system.

{\bf Ba$_2$Nb$_5$O$_x$, BaNbO$_{3-x}$.}
Materials in this class are reported to have T$_c$ as 
high\cite{banbo1,banbo2,banbo3} as high as 22 K.  Values are
strongly sample dependent, but T$_c$ up to 18 K seems to be
reproducible.  The structure that
is suggested to be responsible for the highest T$_c$ is a perovskite
oxynitride BaNbO$_x$N$_y$.  In view of the enormous number of 
non-superconducting perovskite oxides, this could be a particularly
significant achievement if it can be confirmed.

{\bf YPd$_2$B$_2$C.}  This compound is the highest T$_c$ member
(23 K)\cite{cava,chu} of the class of mostly Ni compounds\cite{nagarajan} 
that have been studied most
extensively because of the competition between magnetism (due to 
magnetic rare earths in place of Y) and
superconductivity.  This compound may be regarded as containing 
C-B-C trimers that link YPd$_2$ layers.  Indications from a rigid-potential
treatment\cite{wepdjs} suggests the light atoms provide 
a majority of the coupling
strength, but this tentative conclusion needs to be confirmed by
more complete calculations.

\section{The Denouement}
The primary theme of this paper, if one can be claimed, has been (a) to discuss
how MgB$_2$ and a few other UHTS systems succeed in achieving impressive
T$_c$ by breaking old rules, and (b) to give some thought to the possibility
of extending the positive attributes of these new and different
superconductors.  One observation is that MgB$_2$ puts all its reliance
(coupling) on only a small fraction $\sim$3\% of its phonons.  This
works, as far as it goes.  By varying various materials properties,
we find there is no more than 20-30\% to be gained in this type of system by
increasing the raw coupling strength [N(E$_F$)$<I^2>$], also not much
to be gained by starting with a stiffer system, and also that 
changes in the hole-concentration are ineffective for raising T$_c$.
Increasing raw coupling {\it proportionately} with the stiffness of the underlying 
unrenormalized material is a possible avenue for increasing
T$_c$.  However, we are 
reaching the limits of stiff systems at ambient pressure.  Thinking practically, 
application of such hard materials brings additional headaches --
think of the prospect of winding your electromagnet coils with diamond wire.

\begin{figure}[t]
\label{FS_arrows}
\includegraphics[width=0.40\textwidth]{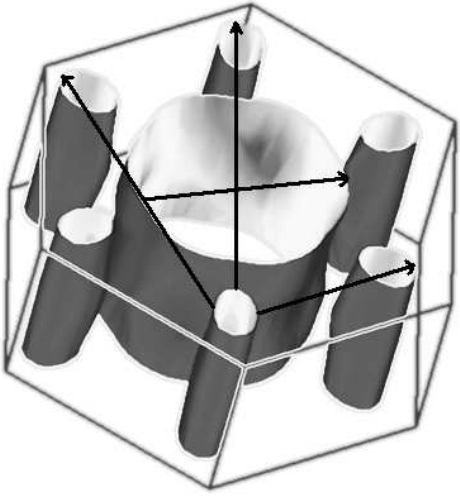}
\caption{Schematic picture of how to enhance total coupling
strength while retaining a stable lattice.  Additional scattering
processes are introduced at large Q in the region of spanning vectors
of Fermi surfaces arrayed around the Brillouin zone.  In this figure 
three spanning vectors are shown, interconnecting the smaller sheets
of Fermi surface. 
}
\end{figure}

This line of pursuit may not quite be hopeless.  Since MgB$_2$ teaches
us that putting lots of eggs into one strong basket is an avenue
to success, then it ought to be the case that putting
even more and bigger eggs into {\it several} strong baskets ought to be
even better.  The mathematical justification is evident: renormalization
of a phonon depends on $\lambda_Q$ for the specific $Q$ point, while
$\lambda$ is a sum over all $Q$.  Whereas increasing coupling strength
in a given region $Q < 2k_F$ reaches its limit (structural
instability) in the way that was
modeled in Sec. II, putting additional strength in other 
(non-overlapping) regions of
the zone $|Q-Q_o| < 2k_F$ adds coupling strength while renormalizing
{\it other} phonons, thus increasing $\lambda$ while not threatening 
the overall stability of the lattice. 

The way one goes about this is illustrated by the Fermi surface in
Fig. 5.  One designs an MgB$_2$-like material ({\it i.e.}
quasi-2D with a stiff reference system) that has {\it several}
cylindrical Fermi surfaces; in the example shown the new feature is
the six (symmetry-related) Fermi surfaces along the $\Gamma$-K 
lines in a hexagonal lattice.  The nesting vectors $Q_n$ (those shown, and
symmetry partners) form the centers of circles (cylinders, when shown
in 3D) of radius $|Q-Q_n|<2k_F$.  These Kohn-anomaly-enclosed
regions have radius $2k_F$, hence diameter of $4k_F$, and
could comprise most of the Brillouin zone, so nearly all
of the phonons are renormalized (and strongly coupled if the bare
coupling is large).  Then, if one is a clever enough materials
designer, one manages to provide a large bare coupling [N(E$_F$)$<I^2>$]
to every branch of the phonon spectrum rather than just two of nine
as in MgB$_2$.  

If one then manages to get all phonons as strongly coupled as in MgB$_2$
(instead of only $\sim$3\%), then one achieves $\lambda \sim$ 25 or
so, with the lattice remaining stable.  For MgB$_2$, the projected
value of T$_c$ (Allen-Dynes equation,\cite{allendynes} 
$\bar \omega$ = 60 meV,
$\mu^*$ = 0.15) is of the order of 400-500 K.  This estimate is in
accord with the stated strong coupling limit 
0.15$\sqrt{\lambda <\omega^2>}$ provided by Allen and 
Dynes,\cite{allendynes} which 
for these constants gives T$_c$ = 525 K.  

This analysis actually neglects the primary aspects of MgB$_2$-like
systems, that 2D phase space is such that the total coupling from a circular
Fermi surface is independent of its size, {\it i.e.} the doping level,
and that the phonon renormalization (and impending structural instability)
also do not depend on the doping level.  Hence it is not strictly the
fraction of the Brillouin zone that one can marshal that is important.
Rather, it is the number of Fermi surfaces one can create -- the
``band degeneracy'' factor $d_B$ in Eq. \ref{lambda}.  Whether
they are hole-like or electron-like is not important, only that they
are there and are quasi-2D.  More, smaller sheets are better, up to a
point; if they get too small (E$_F$ very near a band edge) non-adiabatic
effects arise, and getting into the very low carrier regime will 
introduce a poorly screened Coulomb interaction between carriers that
will invalidate the present considerations. 

Perhaps the most important feature that MgB$_2$ has introduced is a
platform for distributing couplinng strength relatively uniformly,
something that decades ago was presumed to be the norm but may be
instead the exception.  The seemingly innocuous 3D system of fcc Li
achieves a remarkable increase in T$_c$ under pressure, but it arises
from (broadened) `surface regions' in specific locations in the zone.
The correspoinding phonons provide strong coulping but rapidly become
unstable, in line the classical understanding discussed in the Introduction.
Building on ``MgB$_2$-like'' principles might yet lead to important
enhancement of the superconducting critical temperature.

\section{Acknowledgments}
My collaborators on research into the microscopic origins of unusual
electron-phonon coupling in recent years include J. An,
M. D. Johannes, D. Kasinathan, J. Kune\v{s}, A. Lazicki, K.-W. Lee,
H. Rosner, S. Y. Savrasov, and R. Weht.  I have shared several useful
conversations on superconducting materials with O. K. Andersen and
A. Simon, and with many contributors to this volume who attended
the Notre Dame Workshop on {\it The Possibility of Room Temperature
Superconductivity} in June 2005.  I learned much about electron-phonon
coupling from P. B. Allen, who should not however be held responsible 
for my theoretical nearsightedness.
Our recent work in this area has been supported by 
National Science Foundation Grant
DMR-0421810.

\end{document}